\documentclass{aa}
\usepackage{graphicx}
\usepackage{txfonts}
\usepackage{natbib}

\def\kms{$\rm \,km\,s^{-1}$}

\def\H2{H$_2$}

\def\Ha{H$_{\alpha}$}

\def\Pa{P$_{\alpha}$}

\def\Bg{Br${\gamma}$}

\def\micro{$\rm \,\mu m$}

\def\fluxsr{$\rm \,erg\,s^{-1}\,cm^{-2}\,sr^{-1}$}

\def\power{$\rm \,erg\,s^{-1}$}
\def\pc{\,pc}

\def\Xray{X-ray}
\def\Xrays{X-rays}

\begin{document}

   \title{Revisiting the location and environment of the central engine in \object{NGC\,1068}}

   \author{E. Galliano \inst{1}
	\and D. Alloin \inst{1}
	\and G.L. Granato \inst{2}
	\and M. Villar-Mart{\'{\i}}n \inst{3}}

   \institute{European Southern observatory, Casilla 19001, Santiago, Chile
	\and Osservatorio Astronomico di Padova, Padova, Italy
	\and Department of Physical Sciences, University of Hertfordshire, College Lane, Hatfield, Herts AL10 9AB, UK}

   \offprints{E. Galliano}
   \mail{egallian@eso.org}
   

   \date{Received / Accepted }
	\titlerunning{Location and environment of the central engine in NGC\,1068}

   \abstract{
We revisit in this paper the location of the various components observed in the AGN of {NGC\,1068}. Discrepancies between previously published studies are explained, and a new measurement for the absolute location of the K-band emission peak is provided. It is found to be consistent with the position of the central engine as derived by \citet{Gallimore97}, \citet{Capetti97a} and \citet{Kishimoto99}. A series of map overlays is then presented and discussed. Model predictions of dusty tori show that the nuclear unresolved NIR-MIR emission is compatible with a broad range of models: the nuclear SED alone does not strongly constrain the torus geometry, while placing reasonable constraints on its size and thickness. The extended MIR emission observed within the ionizing cone is shown to be well explained by the presence of optically thick dust clouds exposed to the central engine radiation and having a small covering factor. Conversely, a distribution of diffuse dust particles within the ionizing cone is discarded. A simple model for the \H2~and CO emission observed perpendicularly to the axis of the ionizing cone is proposed. We show that a slight tilt between the molecular disc and the Compton thick central absorber naturally reproduces the observed distribution of \H2~of CO emission.
   }

   \maketitle

\section{Introduction}

Efforts in understanding the physics at work in active galactic nuclei (AGN) started in the early seventies, long after the first AGN or quasar had been discovered \citep{Seyfert43,Schmidt59}. Great progress has been made over the last decade, thanks to improved observing tools and modeling explorations. On the side of observations, one can acknowledge the impact of better spatial and spectral resolutions in the near-infrared (NIR), mid-infrared (MIR) and radio wavelengths, as well as the access to high energy photons (UV, \Xrays). On the side of modeling, the availability of complex codes taking into account the large number of physical processes taking place in AGN, and their inter-relations, has permitted to bridge in a more realistic fashion observations and model predictions \citep[see for a review][]{Krolik99}.

It is now commonly accepted that the main source of energy in AGN is of gravitational origin. The various components building up an AGN and its surroundings have also been identified. However, their geometrical arrangement and their links remain to be understood in more detail, with the hope that this will bring also some clues about the formation process of such systems. The commonly accepted AGN model involves a central engine (black hole plus accretion disc, hereafter referred to as CE) surrounded by sets of dense matter clouds (the so-called broad-line and narrow-line regions, hereafter BLR and NLR) and by a dusty/molecular disc-like and thick structure (hereafter the torus) which funnels the emission of high energy photons and particles along privileged directions (the ionizing cone, the radio jet axis). This in turn gives rise to viewing-angle dependent effects which are often invoked to explain the various brands of AGN discovered so far. And last, the AGN, including its close environment, are embedded in the bulge stellar population of the AGN host galaxy.

The intrinsic physical scales of the different components mentioned above are quite different, with orders of magnitude such as: (a) CE (black hole plus accretion disc), $10^{14}$\,cm, (b) BLR, $10^{16}$\,cm, (c) NLR, $10^{20}$\,cm, (d) torus, $10^{18}$ to $10^{20}$\,cm, while the stellar system core hosting the AGN has a typical dimension of $10^{20}$\,cm (but can be more extended). On the other hand, the radio jet and structures can show a huge extension, in some cases on kpc scales. The different AGN components and the components contributing along the line-of-sight close to the AGN (such as the stellar population) can be disentangled in several ways: through their relative flux contributions which vary largely among AGN, possibly reflecting the AGN evolutionary state, and through their different angular sizes, according to the intrinsic sizes mentioned above. However, even in the close-by AGN {NGC\,1068}, which lies at a distance of 14.4\,Mpc, a torus component with intrinsic size $10^{19}$\,cm corresponds to an apparent angular size of 0.04\arcsec. So far, this is out of reach from an observational point of view, except in the radio range with interferometers, or using speckle techniques. 

Therefore, the comparison between observations and model predictions, necessary to probe model parameters, is limited by the angular resolution of available data sets. Conversely, one can focus on physical signatures specific to each component, some of which are listed hereafter. High energy photons and particles arise essentially from the CE (physical processes related to matter accretion and emission at the surface of the accretion disc), while the radio maser emission is thought to be related to the inner walls of the dusty/molecular torus. The flux variations of the CE provide information about the size and structure of the accretion disc. The impact of flux variations on the ionization state of the surrounding BLR clouds makes the BLR line emission respond with a time-delay which reflects the BLR size. The molecular line emission in the NIR to millimeter range, as well as the MIR continuum emission come predominantly from the dusty/molecular component (torus or/and cold material in BLR/NLR clouds shielded from the intense UV radiation field). The contribution from the stellar population starts to fade off at wavelengths above 3.5\micro. In conclusion, a whole set of constraints can be used and this is why the multi-wavelength approach is a very powerful tool in AGN studies. Still, to assemble these pieces of information, we have to face the fact that they have been obtained at different angular resolutions and that they are most often lacking precise astrometric information. These are major sources of limitation which have indeed hampered so far a quantitative probe of model predictions. 

The preferred target on which one could solve the issue is of course the close-by AGN, {NGC\,1068}. Because it is a type 2 AGN, it contains all the components mentioned above; because of its proximity it offers two advantages, its brightness and one of the best intrinsic spatial resolution one can achieve at every wavelength. Thanks to these features, a very large number of observational studies have been performed on {NGC\,1068} since the early seventies (ADS returns 200 entries for refereed articles containing NGC\,1068 in their title, from 1970 to 2002). It is therefore tempting to try assembling major pieces of information obtained so far on this object, with the goal of locating its CE and analyzing the AGN surroundings. 

The first task is to register, on the basis of their absolute coordinates, high resolution maps available at different wavelengths and locate the CE with respect to other AGN components. This is performed in Sect.~\ref{positioning}, where we present a new determination of the absolute positioning of the 2.2\micro~continuum emission peak using ISAAC/VLT data in the \H2~line emission and PdB/IRAM interferometric measurements in the CO(2-1) line and where we compile and discuss the few data sets with astrometric positioning available so far. After having secured the best possible registration, we compare maps obtained at different wavelengths and on different scales, in order to extract quantitative information to be compared with model predictions. 
In Sect.~\ref{SED model}, we discuss the spectral energy distribution (SED) of the unresolved AGN component (intrinsic size less than 8pc FWHM), while the origin of the MIR dust emission observed within the ionizing cone is investigated in Sect.~\ref{extended MIR model}. Finally, we discuss and model the excitation of the molecular material located in the plane perpendicular to the ionizing cone axis (the torus plane) in Sect.~\ref{H2 model}. Concluding remarks and prospective ideas are presented in Sect.~\ref{conclusion}.

\section{Relative positioning of the emission peaks at different wavelengths, location of the CE}
\label{positioning}
The precise location of the CE in the AGN of {NGC\,1068} is still subject to a hot debate. We review hereafter a number of published astrometric measurements at different wavelengths. Then, we present a new absolute positioning of the unresolved K-band continuum peak and we discuss the location of emission peaks at different wavelengths with respect to the CE.

\subsection{The central engine} 

It is commonly admitted that radio observations can directly probe the CE. The 6\,cm MERLIN map of the AGN shows a linear radio structure made of four compact components \citep{Muxlow96,Gallimore96a,Gallimore96c}. One of these components (source S1) is now identified with little doubt as the CE, because of its thermal inverted spectrum between 1.3\,cm and 6 \,cm \citep{Gallimore96c}. Moreover, a disc of H$_2$O and OH maser emission, thought to be associated with the inner walls of a molecular torus around the CE, is observed with the VLBA at the same location \citep{Gallimore96b}. The absolute astrometric precision of the MERLIN images is 20\,mas. \citet{Muxlow96} report the position of S1 at $\rm \alpha=02^h42^m40.7098^s,~\delta=-00^{\circ}00\arcmin47.938\arcsec$ (J2000). In the following, we consider that S1 is the CE.

\subsection{Astrometric positionings}
\label{astrometric positioning}

The objective way to register any map to the 6\,cm MERLIN image is by performing absolute astrometry. Any other method is subjective in the sense that it will require some physical assumption.

Even though 0.1\arcsec~resolution HST UV-continuum and optical NLR images have been published before 1997 \citep{Evans91, Macchetto94}, they could not be precisely registered since the absolute positioning of the HST images was only possible with a precision of $\pm$0.5\arcsec. Based on a method developed by \citet{Lattanzi97}, \citet{Capetti97a} report for the first time on an absolute registration of HST images with an $\pm$0.08\arcsec~precision. They find that the optical (F547M filter) continuum peak (hereafter referred to as HST F547M continuum) is located at $\rm \alpha=02^h42^m40.711^s,~\delta=-00^{\circ}00\arcmin47.81\arcsec$ (J2000), that is  0.13\arcsec$\pm$0.1\arcsec~North and 0.02\arcsec$\pm$0.1\arcsec~East of S1. This corresponds to the peak of the NLR-cloud B. They also show that the UV (F218W) and F547M peaks are coincident within 25\,mas.

\begin{figure*}[htbp]
\begin{center}
\resizebox{18cm}{!}{\includegraphics*[scale=1.]{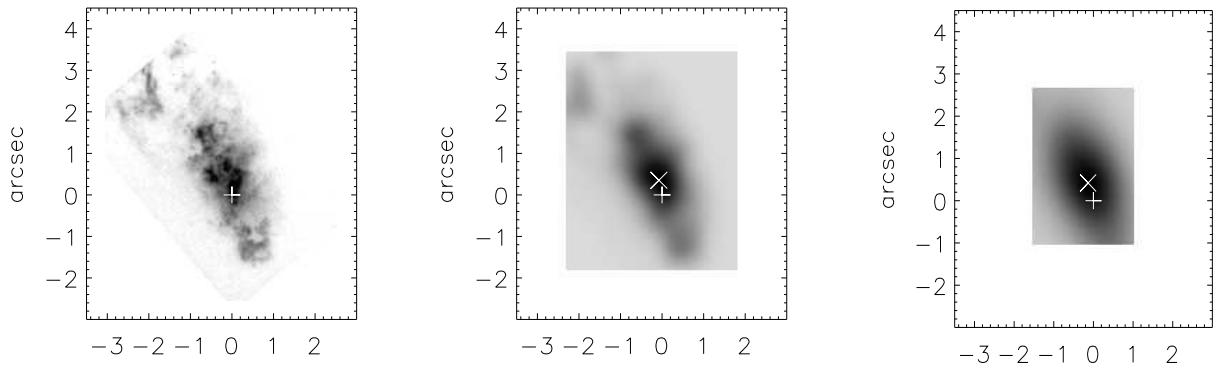}}
\caption{Illustration of the seeing effect on the [OIII] image of the ionizing cone in NGC\,1068. From left to right: HST [OIII] image, same image convolved with a 0.5\arcsec~Gaussian, same image convolved with a 1.0\arcsec~Gaussian. The ``+'' marks the position of the HST [OIII] peak and the ``$\times$'' marks the position of the [OIII] peak in the images degraded by seeing effects.}
\label{OIII convolve}
\end{center}
\end{figure*}

On the photographic 103a-O plates they use to perform the absolute astrometry, \citet{Capetti97a} locate the visible photocenter at $\rm \alpha=02^h42^m40.727^s,~\delta=-00^{\circ}00\arcmin47.69\arcsec$ (J2000). The difference between the locations of the HST F547M continuum peak and the photographic plate visible photocenter is 0.27\arcsec. Such a large difference can arise from two effects. First the F547M filter cuts out the prominent [OIII] $\lambda$5007 emission lines, while the ground-based 103a-O photographic image does sense these lines, the photocenter of which may be slightly displaced from the continuum photocenter. However, such a displacement of the [OIII] peak with respect to the F547M peak only occurs at the scale of individual NLR clouds. Indeed \citet{Macchetto94} show that the UV continuum emission peak in NLR-cloud B is shifted by 0.05\arcsec~towards the North with respect to the [OIII] peak. Actually, the main reason for the observed shift of 0.27\arcsec~is that the photographic image, which has a low spatial resolution (larger than 1\arcsec), mixes the different continuum and [OIII] emitting clouds. Indeed, one observes that the photographic plate photocenter is shifted towards the North-East with respect to the F547M peak. This is consistent with the overall shape of the NLR, which is extended towards the North-East. Considering the complexity of the NLR visible continuum and [OIII] emission, images at different spatial resolutions ought to produce emission peaks at different locations. 
This effect must be considered carefully when interpreting the relative astrometric measurements between the emission peaks in the optical and in the IR, since the bulk of the NIR emission is unresolved down to 0.03\arcsec~\citep{Wittkowski98}, while the optical emission is made of HST-resolved clouds over a region of $\sim$0.5\arcsec~$\times$1\arcsec. Fig.\,\ref{OIII convolve} illustrates this effect by showing the 0.1\arcsec~HST [OIII] image taken from \citet{Macchetto94}, and the same image after convolution with 0.5\arcsec~and 1.0\arcsec~FWHM Gaussian profiles, simulating seeing effects. The shifts of the position of the [OIII] peak on the images respectively degraded by an 0.5\arcsec~and 1.0\arcsec~seeing effect are 0.35\arcsec~and 0.45\arcsec, with respect to the HST original image. This is of the same order of magnitude as the measured offset between the HST F547M peak and the 103a-O visible photocenter, sensing the continuum (and [OIII] lines for the 103a-O plates). As a consequence, the results of \citet{Braatz93} and \citet{Marco97} which position the N-band peak and K-band peak at 0.3\arcsec~South, 0.1\arcsec~West from respectively the optical peak (from an optical TV camera) and the I-band peak should be considered with caution: they both positioned the MIR or the NIR peaks with respect to the visible peak of a {\bf low resolution image}. In order to gain in precision, we discuss in the following the positioning of the NIR peak with respect to the optical continuum peak of NLR-cloud B only.   

\subsection{Assumption-depending positionings}
\label{physical positioning}
Positioning of the emission peaks at different wavelengths with respect to the CE have also been performed under various physical assumptions. 
\subsubsection{Positioning of the K-band peak}

The absolute astrometry of the K-band emission peak can be performed directly in two ways. 

\begin{figure*}[htbp]
\begin{center}
\caption{\textbf{(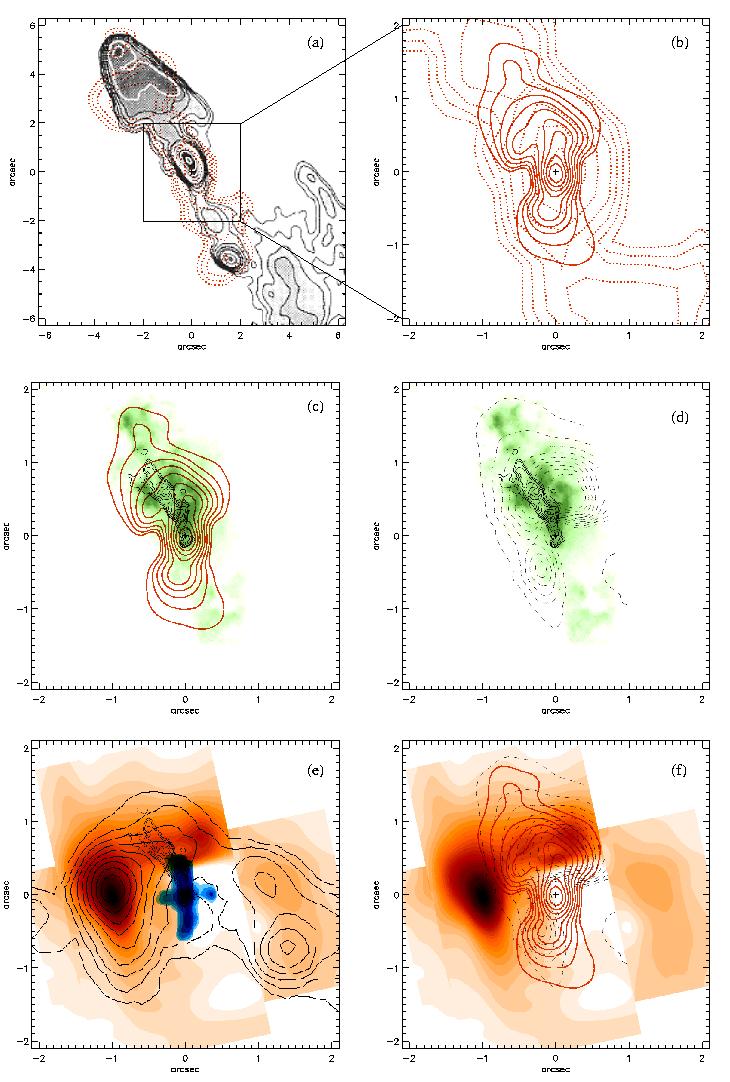)} Map superpositions. Axis unit are arcsec, North is up and East to the left. Coordinates (0\arcsec;0\arcsec) give the location of the CE (see Sect.\,\ref{positioning}). From top to bottom, left to right: (a) 12\arcsec$\times$12\arcsec~region around the CE of {NGC\,1068}. The grey contours show the 6cm emission \citep{Gallimore96a}; the red dotted contours show the 20\micro~image \citep{Alloin00}. (b) Zoom to the central 4\arcsec$\times$4\arcsec~region. Dotted red contours are the same contours as in map (a); the solid red contours show the 12.3\micro~map \citep{Tomono01}. (c) In green is displayed the [OIII] image by \citet{Macchetto94}; in black contours the 6\,cm map by \citet{Gallimore96a}; red contours are the same as in (b). (d) The dashed contours correspond to the \Bg~image by \citet{Galliano02}; the green image and the black thin contours are as in (c). (e) The orange map shows the \H2~2.12\micro~emission from \citet{Galliano02}; the large scale contours show the $^{12}$CO(2-1) map from \citet{Schinnerer00}; the small scale contours are the same as in map (c); the blue map displays the M-band image by \citet{Marco00} (f) The orange image is as in (e); the red contours as in (b); the dashed contours as in (d)}
\label{superpositions maps}
\end{center}
\end{figure*}

The first method is to compare the high resolution CO emission line map, for which absolute astrometry is available, and the \H2~2.12\micro~emission line map, which appears to be very similar to the CO map and for which there is a relative position of the K-band peak. \citet{Tacconi97} and \citet{Schinnerer00} observed at $\sim$0.5\arcsec~resolution the CO(2-1) line emission in the millimeter wavelength range with the IRAM interferometer. They produced a CO map of the central 3\arcsec~ region, finding that the CO emission is asymmetrically distributed, with a strong CO emission knot to the East of the CE. The CO emission peak corresponding to this knot is located at $\rm \alpha=2^h42^m40.767^s$ and $\rm \delta=- 0^{\circ}00\arcmin47.95\arcsec$ (J2000), with a precision of $\pm$0.05\arcsec. In parallel, \citet{Galliano02} produced a map at similar resolution of the \H2~2.12\micro~emission derived from VLT/ISAAC observations. The CO and \H2~maps correlate very well (Fig.~\ref{superpositions maps},e). Measurements on the data presented in \citet{Galliano02} show that the eastern \H2~knot is located 0.95\arcsec$\pm0.15\arcsec$ East and 0.1\arcsec~$\pm$0.15\arcsec~ North of the K-band continuum peak. The similarity of the two maps and the fact that, in the AGN context, CO and \H2~emission are expected to be coincident \citep{Maloney96}, allow us to assume that the CO and \H2~peaks are at the same location within 0.1\arcsec. The absolute coordinates of the K-band peak can therefore be derived: $\rm \alpha=2^h42^m40.702^s \pm 0.012^s$ and $\rm \delta=-00^{\circ}00\arcmin48.04\arcsec \pm 0.18\arcsec$ (J2000). This position of the K-band emission peak is consistent with the position of S1, while it is only marginally consistent with the positions of the HST F547M continuum and the [OIII] peaks.

Another positioning of the K-band peak has been performed by \citet{Thompson01}. Using the similarity of the NLR features seen in the HST/WFPC2 \Ha~image and in the HST/NICMOS \Pa~image and using the fact that both \Ha~and \Pa~are Hydrogen recombination lines, and consequently should trace the same features, they have registered the position of the NIR continuum image with respect to the HST F547M continuum image. They find the location of the K-band peak to be consistent with both the center of the UV polarization pattern as determined by \citet[ see Sect.\,\ref{physical positioning}]{Kishimoto99}, and the [OIII] peak (NLR-cloud B). This coincidence is claimed to be accurate to $\pm0.04\arcsec$.     
\begin{table*}[htbp]
\caption[]{Absolute positions of the distinct components of the nuclear region of {NGC\,1068}}
\begin{center}
\begin{tabular}{llll} \hline \\[-0.3cm]
 & RA (J2000) & dec (J2000) & reference\\ \hline
F547M peak & $\rm 02^h42^m40.711^s \pm 0.005^s$ & $\rm -00^{\circ}00\arcmin47.81\arcsec \pm 0.08\arcsec$ & \citet{Capetti97a}\\
S1 & $\rm 02^h42^m40.7098^s \pm 0.001^s$ & $\rm -00^{\circ}00\arcmin47.938\arcsec \pm 0.02\arcsec$ &\citet{Muxlow96}\\
K peak &  $\rm 02^h42^m40.702^s \pm 0.012^s$ & $\rm -00^{\circ}00\arcmin48.04\arcsec \pm 0.18\arcsec$ & this paper\\
UV polarization &  $\rm 02^h42^m40.713^s \pm 0.006^s$ & $\rm -00^{\circ}00\arcmin47.93\arcsec\,_{-0.11\arcsec}^{+0.14\arcsec}$ & \citet{Kishimoto99}\\[-0.15cm]
pattern center &  & &\\
\hline\\[-1cm]
\label{table positions} 
\end{tabular}
\end{center} 
\end{table*}

\subsubsection{Positioning of the center of the polarization patterns in the UV and in the IR}

\citet{Capetti95a,Capetti95b} have presented HST imaging polarimetry of the NLR in the UV (F218W) and optical (F555W). The polarization vectors are found to be distributed in a circular pattern, as expected in the case of scattered radiation from a point-like source, assumed to be the CE. After a correction of their first result \citep{Capetti95a}, they claimed to have determined the position of the center of polarization, at the location 0.45\arcsec~$\pm$0.015\arcsec~South, 0.06$\pm$0.015\arcsec~West of the optical (F555W) continuum peak. However, \citet{Kishimoto99} noticed that in the map of polarization position angles (PA) of \citet{Capetti95a,Capetti95b}, clear deviations from the centrosymmetric pattern can be seen, although not discussed in that paper. Therefore, \citet{Kishimoto99} has extended this work, performing a very complete error budget, and has redetermined the location of the symmetry center of the polarization pattern in the UV within a convincing error circle. He finds a location for the CE which is significantly different from that of \citet{Capetti95b}. The revised location is at $\rm \alpha\,=\,02^h42^m40.713^s \pm 0.0006^s$ and $\rm \delta\,=\,-00^{\circ}00\arcmin47.93\arcsec\,_{-0.11\arcsec}^{+0.14\arcsec}$ (J2000).

\citet{Lumsden99} have performed NIR and MIR imaging polarimetry at good spatial resolution ($\sim$0.5\arcsec~ in the NIR). They detect NIR polarized emission in extended regions along PA$\sim$30\degr~both towards the North and the South. In the optical the light from the southern part is absorbed, probably by matter in the plane of the host galaxy. They suspect that an additional mechanism, other than scattering, is contributing to the polarization at all wavebands close to the AGN. Therefore, they have masked the central 2\arcsec~region and derived the scattering center from the outer regions only. They find offsets between the polarization pattern center and the flux centroid in the three NIR bands (J,H,K) of respectively 0.11\arcsec~$\pm$0.18\arcsec~, 0.09\arcsec~$\pm$0.21\arcsec~ and 0.34\arcsec~$\pm$0.39\arcsec~, offsets which are formally consistent with no offset. Unfortunately, they do not provide the PA of the offsets. If the PAs for the three offsets are different, then their measurements will actually be consistent with no offset, while on the contrary, if the three offsets were for example at PA$\sim$30\degr, then their result would be consistent with the UV polarization center found by \citet{Kishimoto99}. 

Recently, \citet{Simpson02} have published NIR (1.9-2.1\micro) polarimetry from HST/NICMOS data set. They claim that the center of the NIR polarization pattern does correspond to the NIR peak within $\pm0.08$\arcsec. \\

\begin{figure}[htbp]
\begin{center}
\resizebox{9cm}{!}{\includegraphics*[scale=1.]{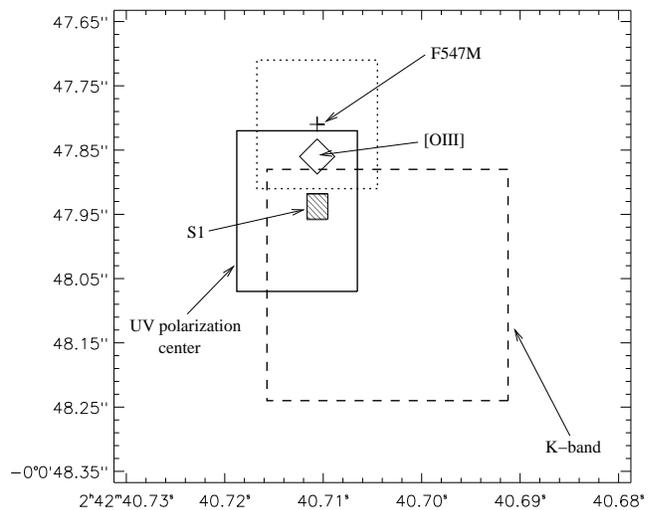}}
\caption{Positions of the different peak, as described in Sect.\,\ref{positioning} on the J2000 absolute coordinate reference frame. The errors represented by the boxes are given with respect to the absolute coordinate reference frame. The hatched rectangle marks the location of S1 (CE). 
The dotted line rectangle corresponds to the location of the F547M continuum peak. The cross marks its center, while the diamond gives the {\bf relative} position of the [OIII] peak (NLR-cloud B) with respect to the F547M peak; the size of the diamond gives the error of this relative measurement. The plain line rectangle corresponds to the center of the UV polarization pattern center \citep{Kishimoto99}. The dashed line rectangle represents our new measurement of the K-band peak location.}
\label{positions1}
\end{center}
\end{figure} 
\begin{figure}[htbp]
\begin{center}
\caption{\textbf{(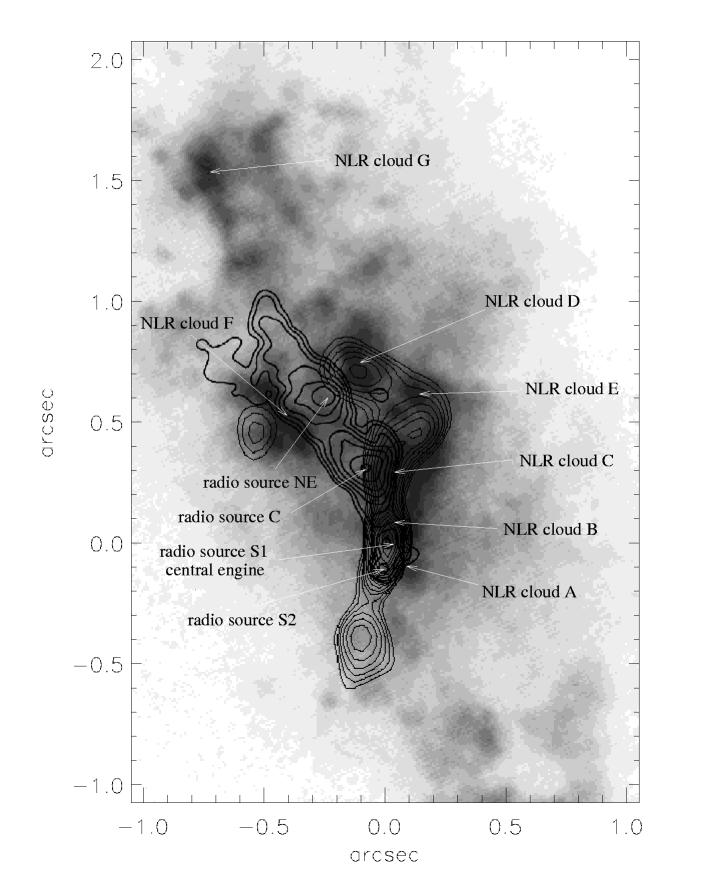)} Superposition of the [OIII] image of the NLR by \citet{Macchetto94} (grey scale), MERLIN 6cm image \citep[thick contours][]{Muxlow96}, and 12.7\micro~image by \citet{Bock00} in thin contours, following the positions given in table~\ref{table positions}. The labels are from \citet{Evans91} for the NLR clouds and from \citet{Gallimore96a} for the radio sources. Coordinates (0\arcsec,0\arcsec) are assigned to S1.}
\label{superposition1}
\end{center}
\end{figure} 
\subsection{Summary and conclusion}
\label{summary}
We have compiled measured positions of the emission peaks at different wavelengths, of the centers of the polarization patterns in the UV and in the IR and of the positions of the radio sources in the region surrounding the AGN in {NGC\,1068}. We find that the measurements converge to a unique picture, and that all the components are relatively positioned with a good precision. A sketch of the positions of the most important features is displayed in Fig.\,\ref{positions1} while their absolute coordinates are listed in Table~\ref{table positions}. 

\citet{Thompson01} found that the K-band peak is coincident with the position of the UV polarization pattern center and the peak of [OIII]$\lambda$5007 in the NLR-cloud B, at $\pm0.04\arcsec$ precision. \citet{Simpson02} found that, within $\pm$0.08\arcsec, the NIR (1.9-2.1\micro) polarization pattern is centered on the NIR peak. These three measurements converge to position the NIR peak at the position of S1 within $\pm$0.1\arcsec. We do not contemplate the offsets between visible peaks and N- and K-band peaks measured respectively by \citet{Braatz93} and \citet{Marco97}, since these offsets are affected by the use of images at different spatial resolutions (Sect.~\ref{astrometric positioning}). 
Regarding the UV polarization pattern center, we consider only the analysis of \citet{Kishimoto99}, which is the most precise. 
There is no absolute positioning of the MIR maps available. A very reasonable assumption, supported by dusty torus models \citep{Granato94,Granato97}, is to consider that the MIR emission peak appears closer to the CE than the NIR emission peak. The reason for this is that the optical depth in the NIR is larger than in the MIR and the scattering efficiency is much higher in the NIR than in the MIR. Therefore, the NIR photons that we see might well be photons which were scattered by the dust along the axis of the torus, where the optical depth to the observer is smaller, and consequently, the center of light in the NIR may be displaced towards the scattering region, away from the CE. As the wavelength increases in the MIR, the optical depth to the CE decreases, as does the scattering efficiency, and the MIR light peak gets closer to the location of the CE.  

Fig.\,\ref{superposition1} displays a superposition of the radio 6cm contours and the MIR 12.7\micro~contours over an image of the NLR in the [OIII]$\lambda$5007 line emission. The positions are set according to Sect.\,\ref{summary}. The coincidence found between the NLR [OIII] emitting clouds and the MIR emission suggests that the same physical clouds are responsible for both emissions. With this registration, the radio source C is not coincident with the NLR-cloud B, as suggested in the preferred registration of \citet{Gallimore96a}. If one attempts to identify a particular NLR cloud responsible for the deflection of the radio jet, it would more likely be NLR-cloud C. The fact that NLR-cloud B does not affect the trajectory of the jet suggests that it lies on the foreground (or background) of the jet. The anti-correlation between the [OIII]+MIR images on one hand and the radio map on the other hand, is striking. It looks as if the [OIII] and MIR light were emitted principally in regions surrounding a cavity carved by the radio jet and localized at the edges of the ionizing cone. The matter inside the cone may have been swept away by the jet, and pushed to the actual location of NLR-cloud G, 1.7\arcsec~from S1 in the direction of the deflected radio jet. \citet{Gallimore96a} originally proposed that the gaps in the NLR clouds may be caused by the jet, guessing an alignement between the MERLIN and the HST images, that turns out to be close to the actual alignement, as found in our analysis.    

\subsection{Map superpositions}
\label{superpositions}

Considering the absolute and relative positionings discussed above, Fig.\,\ref{superpositions maps} presents a series of superpositions of maps at different wavelengths and on different scale. The caption describes each map in detail. 

Map (a) shows the excellent correlation between the {\bf large scale} 6cm map and the 20\micro~image. At this resolution (0.6\arcsec), each 20\micro~peak has a counterpart on the 6cm map. 

Map (b) reveals the good correlation existing between the low resolution (0.6\arcsec) 20\micro~image and the high resolution (0.1\arcsec) 12.3\micro~map. This demonstrates the very good consistency of the two datasets.

Map (c) shows the correlation between the [OIII] and the MIR emission and the anti-correlation between the [OIII] and the 6\,cm radio emission on {\bf small scale}. This is a strong evidence in favor of ionized clouds containing large amounts of cool dust. 

Map (d) indicates that, as expected, the \Bg~emission is correlated with the [OIII] emission. It also shows that the \Bg~emission is strongly enhanced at the location of NLR-cloud E. Its particular excitation, the fact that it is not a strong MIR emitter (see Fig.\,\ref{superposition1}) and its location with respect to the radio jet suggest that NLR-cloud E is undergoing a different excitation than the other NLR-clouds. 

Map (e) highlights the very good correlation between the \H2~2.12\micro~and the $^{12}$CO(2-1) maps. Still one notices the inverted brightness ratios in CO and \H2~in the two western knots. On the contrary the eastern knot is very bright in both lines. This map also shows the spatial coincidence, at a 0.5\arcsec~scale, between the radio emission and the M-band emission. One even notices that the M-band emission bends together with the radio emission, at about 0.3\arcsec~North of the nucleus.

Map (f) summarizes the locations of the \H2~2.12\micro~and \Bg~line emission, and the 12.3\micro~emission.

\section{The unresolved torus emission}
\label{SED model}
\begin{figure}[h]
\begin{center}
\resizebox{9cm}{!}{\includegraphics*[scale=1.]{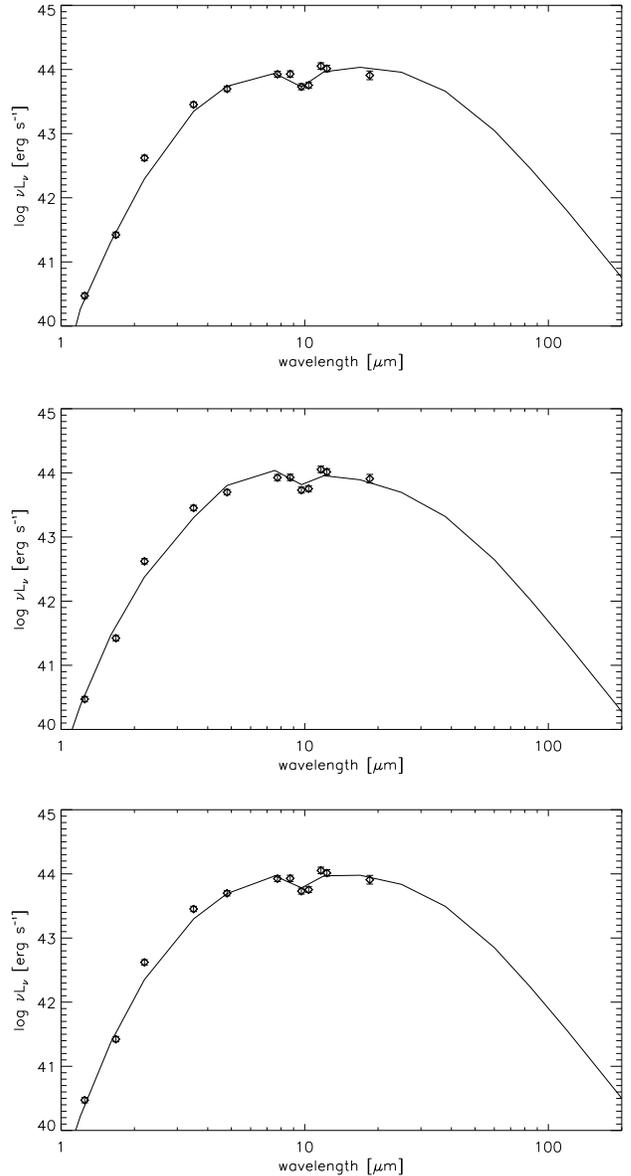}}
\caption{Examples of fits for tapered disc models. 
The model prediction is shown as a continuous line: from top to bottom, the model parameters corresponds to models 1,2 and 3 in Table~\ref{param values}. The observed SED is represented by diamonds (plus error bar) and corresponds to the fluxes displayed in Table~\ref{n1068 fluxes}.}
\label{fitnuctap}
\end{center}
\end{figure}


The observed IR SED of {NGC\,1068} and other AGN have been compared to predictions from radiative transfer models within dusty tori, in order to investigate the validity of the unified model and constrain possible geometries for the obscuring structure \citep[e.g.][]{Pier93,Granato94,Efstathiou95,Granato97,Alonso-Herrero01,Nenkova02}. A major problem is that it is still quite difficult to assess the \textbf{truly nuclear} SED to be compared with torus models, even for the best studied AGN, {NGC\,1068}. In the NIR bands, even small aperture photometry is likely to include a non-negligible contribution from starlight. For this object, the best option so far is to combine the nuclear fluxes by \citet{Rouan98} in the J-,H-,K-bands (corrected for star contamination) and by \citet{Marco00} in the L-,M-bands with the nuclear MIR fluxes provided in \citet{Tomono01} since they are all derived from high spatial resolution images. The resulting nuclear SED is shown on Fig.~\ref{fitnuctap} and the flux values reported in Table.~\ref{n1068 fluxes}.

\begin{table}[htbp]
\caption[]{IR fluxes for {NGC\,1068}. The references are the following: R98=\citep{Rouan98}, M00=\citep{Marco00}, T01=\citep{Tomono01}}
\begin{center}
\begin{tabular}{ccccc} \hline \\[-0.3cm]
Wavelength & Band   & Flux & Aperture \diameter& Reference\\
\micro    &        & Jy & arcsec &           \\ \hline
1.25 & $J$ & $5\,10^{-4}\pm10\%$ & 0.2 & R98\\
1.7  & $H$ & $6\,10^{-3}\pm10\%$ & 0.2 & R98\\
2.2  & $K$ & $0.124\pm10\%$ & 0.2 &      R98\\
3.5  & $L$ & $1.34\pm10\%$ & 0.4 & M00\\
4.8  & $M$ & $3.23\pm10\%$ & 0.4 & M00\\
7.7  & $N$ & $8.8\pm1.0$ & 0.4 & T01\\
8.7  & $N$ & $10.0\pm1.2$ & 0.4 & T01\\
9.7  & $N$ & $7.04\pm0.8$ & 0.4 & T01\\
10.4 & $N$ & $7.94\pm0.87$ & 0.4 & T01\\
11.7 & $N$ & $17.8\pm2.2$ & 0.4 & T01\\
12.3 & $N$ & $17.2 \pm2.1$ & 0.4 & T01\\
18.5 & $Q$ & $20.2\pm3.4$ & 0.4 & T01\\
\hline\\[-1cm]
\label{n1068 fluxes} 
\end{tabular}
\end{center} 
\end{table}

In the absence of precise physical ideas concerning the structure of the obscuring torus, several geometries -- and associated free-parameters -- are plausible and indeed have been investigated in the papers quoted above (flared discs, tapered discs, cylinders, with and without substantial clumping etc..).

An usually overlooked problem in this kind of study, is that dust optical properties around AGN are likely to be to some extent peculiar. As a matter of fact, \citet{Maiolino01a,Maiolino01b} presented evidence for 'anomalous' properties of dust in AGN. Here the term 'anomalous' is relative to the standard properties of dust grains thought to be responsible for the average Milky Way extinction law and cirrus emission. As pointed out by \citet{Maiolino01a,Maiolino01b}, it is not surprising that the properties of dust grains might be very different  in the dense and extreme environment of an AGN. In particular, they invoke a dust distribution biased in favor of large grains, ranging up to 1-10 $\mu$m, much larger than the usual cut at $\sim 0.3 \mu$m of \citet{Mathis77} type models \citep[e.g.][]{Draine84,Silva98}. This solution has been introduced as the best, among the many explored, to explain the well established decrease of $\rm E_{B-V}/N_H$ and $\rm A_V/N_H$ ratios in the line-of-sight of AGN. It is worth noticing that the tendency for larger $R_V \equiv A_V/(A_B-A_V)$ found in dense galactic regions are commonly explained in a similar way \citep[ and references therein]{Draine01,Maiolino02}.

Thus, we have performed SED fitting, adopting several different geometries, but also investigated models in which the size distribution of grains extends to radii, $a_{max}$, larger than the standard value. Of course, the uncertainty on optical properties of dust further limits the already moderate success of this approach in constraining the geometry (see below).

Due to the relatively long computing time required to produce a model, the SED fitting is performed through comparison with libraries of models. Each library consists of several hundreds of models belonging to a given "geometry class", in which typically 4-5 parameters are varied assigning to them 3-4 different values over a quite wide range. In particular we have compared the NGC\,1068 nuclear SED with both anisotropic flared discs and tapered discs \citep[for definitions, see ][]{Efstathiou95}. The former geometry consists of a structure whose height above the equatorial plane $h$ increases linearly with the distance in the equatorial plane $r$ \citep[Fig.\,1a in ][]{Efstathiou95}. Therefore the dust-free region (ignoring here dust possibly surviving in the NLR clouds) is exactly conical, with half-opening angle $\Theta_h$. We also introduce \citep[at variance with respect to ][]{Efstathiou95} a dependence of the dust density $\rho$, and therefore of the radial optical depth, on the polar angle $\Theta$. The general form used for this dependence and that on the spherical coordinate $r$ is:

$$
\rho(r,\Theta) \propto r^{-\beta}  \exp(-\alpha\,
\cos^2 \Theta)
$$
where $\alpha$ and $\beta$ are model parameters.

In tapered discs, $h$ increases with $r$ only up to a max value $h_{\rm max}$. In this case the dust density within the disc has been always kept independent of $\Theta$. The two additional parameters used to fully characterize the models in both cases are the ratio between the outer and inner (i.e. sublimation) radii $r_{max}/r_{min}$, and the equatorial optical depth to the nucleus at $9.7 \mu$m,  $\tau_{e,9.7}$

The parameter values used in  the libraries are:
\begin{itemize}
\item for both flared and tapered discs:
$$a_{max}=0.3, 0.5, 1.0 \, \mu\mbox{m}$$
$$\beta=-1, 0, 0.5, 1, 2$$
$$\tau_{e,9.7}=0.1, 0.3, 1, 3, 10$$
$$r_{max}/r_{min}=30, 100, 300$$
\item for flared discs only:
$$\Theta_{h}=20,30,40 \, \mbox{\degr}$$
$$\alpha=0, 1,  3, 6$$
\item for tapered discs only:
$$\Theta_{h}=30 \, \mbox{\degr}$$
$$h_{max}/r_{max}=0.2,0.3,0.4$$
\end{itemize}

The impacts of changing various parameters on the predicted SED have been explored in \citet{Granato94} and \citet{Granato97}. Here we recall only that, roughly speaking, $r_{max}/r_{min}$ is related to the width of the IR bump, whilst $\tau_{e,9.7}$ controls mainly the NIR slope of the SEDs, as observed from obscured directions, as well as its anisotropy.

All possible combinations of parameter values are considered in the libraries. For each corresponding model we compute the SED for viewing angles from 0\degr~to 90\degr~in steps of 10 degrees. Then we search through these libraries which models provide a reasonable fit to the observed SED. The employed merit function is $\chi^2$. We have assigned to each data point an error of 30\%. This could well be an underestimate of the true uncertainty of the nuclear flux, especially in NIR bands, due to the difficulty of disentangling the starlight contribution. For instance, the nuclear K-band flux (corrected for stellar light) by \citet{Alonso-Herrero01} in 3\arcsec~aperture is greater than our value by as much as a factor 3. 

Indeed, the main result of our analysis is that several rather different combinations of geometry and parameters of the torus are perfectly compatible with the observed IR SED of  NGC\,1068 (examples of library fits are shown in Fig.\,\ref{fitnuctap} for the tapered models). In other words, as expected, the nuclear SED alone does not allow to put strong constraints on geometrical parameters. Several previous analysis have been too optimistic on this aspect, due essentially to an insufficient exploration of the parameter space. There is some tendency though, for the models with $a_{max}$ larger than the standard value to allow more combinations of the other parameters and still get an acceptable fit. We also confirm that some constraint can be put on the torus extension and on the optical thickness, and we find that good fits are obtained both with tapered (fits shown in Fig.\,\ref{fitnuctap}) and flared disc models, provided the tori are moderately thick, in the sense discussed by \citet{Granato97}. The development of interferometric observations should, in the future, provide more observational constraints. %

\begin{table}[htbp]
\caption[]{SED fitting parameters. The parameters meaning is given in the text. ``t'' refers to tapered discs, and ``f'' to flared discs.}
\begin{center}
\begin{tabular}{lllllll} \hline \\[-0.3cm]
fit number & 1 & 2 & 3 & 4 & 5 & 6 \\ \hline
disc type  & t & t & t & f & f & f\\
$a_{max}$ & 0.3 & 0.5 & 0.5 & 1.0 & 1.0 & 0.3 \\
$\tau_{e,9.7}$ & 10 & 10 & 10 & 10 & 10 & 3 \\
$i$ & 40 & 50 & 40 & 30 & 30 & 30\\
$r_{max}/r_{min}$ & 100 & 100 & 100 & 30 & 30 & 100\\
$\Theta_h$ & 30 & 30 & 30 & 20 & 20 & 20\\
$\alpha$ & - & - & - & 0 & 0 & 0 \\
$h_{max}/r_{max}$=& 0.4 & 0.4 & 0.2 & - & - & -\\ 
$\beta$ & 0.5 & 0.5 & 0.5 & 0.0 & 0.5 & 0.5 \\

\hline\\[-1cm]
\label{param values} 
\end{tabular}
\end{center} 
\end{table}

\section{The extended MIR emission associated with the ionizing cone.}
\begin{figure*}[htbp]
\begin{center}
\resizebox{15cm}{!}{\includegraphics*[scale=1.]{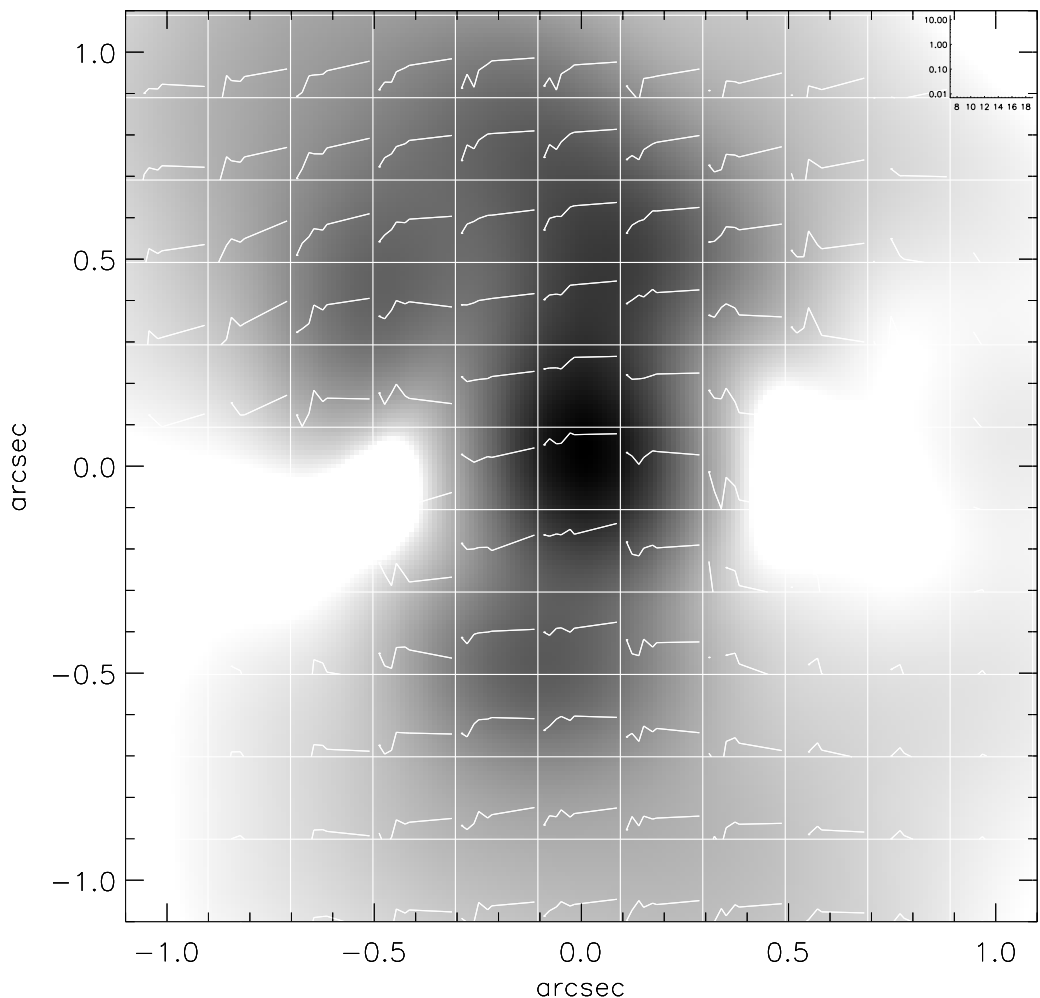}}
\caption{Background image: the 12.3\micro~map from \citet{Tomono01}. In each cell of the image, we show the observed local SED between 7\micro~and 20\micro. Each SED is the total SED of the corresponding cell. The units for the SEDs are Jy for the y-axis and \micro~for the x-axis, with the scales given in the insert on the top-right.} 
\label{mapminised}
\end{center}
\end{figure*}
\label{extended MIR model}
MIR high resolution images of {NGC\,1068} show elongation perpendicularly to the putative dusty torus-like structure \citep{Alloin00,Bock00,Tomono01}. This is interesting in two ways: (i) the extension of the emission, of the order of 70\,pc, exceeds the estimated size of the torus, which is not expected to be larger than 20\,pc \citep{Granato94,Efstathiou95,Granato97}, and (ii) it traces approximately the two ionization cones where dust grains are less likely to survive. Also, this emission is more prominent on the North-East side of the CE (which is believed to trace the cone closer to our line of sight) than on the opposite South-West side. The decline of the surface brightness with distance from the CE is relatively shallow and the observed "local" SEDs of the emission (see Fig.\,\ref{mapminised}) show interesting trends: in the North-East side, the SEDs along the axis have a steep decline towards shorter wavelengths, while in the South-West side, the SEDs appear to be flatter.

\subsection{Diffuse dust in the cone?}
Is it possible to explain the observations with \textbf{diffuse} dust distributed in the cone? To investigate this hypothesis, we have run several sets of models with diffuse dust in the cone extending up to $\sim 1$\arcsec, with different density and grain size distributions, and could not find a good match with the observations. Such models are affected by either, but usually both, of the following problems: (i) the emitted specific intensity decreases more steeply than the observed MIR maps show, and/or (ii) the local SEDs in the cones are too flat. The dust volume emissivity can be enhanced either by increasing the dust density or its temperature. The first possibility works only up to a certain point above which the dust in the cone becomes optically thick to optical-UV photons and, as a consequence, is less heated by the CE. We have tested with our model that this ``turnover'' is reached before the observed emission can be produced (see also the analytical estimate below). Moreover, an excess density of diffuse dust in the cone would obscure the CE and the BLR from any direction, and would also require a different mechanism than direct photo-ionization from the CE to produce the conical NLR.

Let us perform as well an analytical calculation in order to evaluate when this ``turnover'' is reached. Let us assume a distribution of homogeneous diffuse dust in the cone, optically thin in IR (not necessarily in optical-UV). The observed specific intensity $I_{\nu,18}$ at 18 $\mu$m is:
$$
I_{\nu,18}=l \, \alpha_{\nu,18} B_{\nu,18}(T)=\tau_{\nu,18}(l) \,
B_{\nu,18}(T)
$$
$\tau_{\nu,18}(l)$ is the optical depth along the line of sight crossing the cone (l), $\alpha_{\nu,18}$ the extinction coefficient and $B_{\nu}$ is the Planck function at the frequency $\nu$. Specific quantities are all evaluated at 18\micro. Given the geometry, we can evaluate the dust temperature $T$ assuming that the heating is only from the CE and that the distance from the central source, $r$, is almost constant along the line-of-sight. Also, $r \simeq l$. Thus we can identify the optical depth $\tau_{\nu,18}(l)$ with the optical depth to the central source $\tau_{\nu,18}(r)$. For graphite grains of typical size, the equilibrium dust temperature $T$ can be estimated as \citep{Barvainis87}:
$$
T=1650 \left(\frac{L_{UV,46}}{r_{pc}^2}\right)^{1/5.6}
\exp-\frac{\tau_{UV}(r)}{5.6}
$$
Setting $r=70$ pc (i.e.\ 1\arcsec, where the observed $I_{\nu,18}$
is $\sim 10$ Jy\,arcsec$^{-2}$) and $L_{UV}=0.02 \times 10^{46}$
\power~we have
$$
T=180 \exp-\frac{\tau_{UV}(r)}{5.6}
$$
On the other hand, for standard Galactic dust mixture,
$\tau_{UV}\simeq 80 \times \tau_{\nu,18}$. This leads to:
$$
I_{\nu,18}=\frac{\tau_{UV}(r)}{80} \, B_{\nu,18}\left(180
\exp-\frac{\tau_{UV}(r)}{5.6}\right)
$$
This function of $\tau_{UV}$ has a maximum of $\sim 10$ Jy\,arcsec$^{-2}$ (i.e.\ more or less the observed level) at $\tau_{UV}\simeq 1$, but as long as the dust is optically thin in the cones at optical-UV wavelengths -- the condition for photoionization -- the predicted 18\micro~brightness is significantly lower than observed (for instance by a factor 4 for $\tau_{UV}\simeq 0.1$, or by a factor 100 for $\tau_{UV}\simeq 0.01$).

To increase the dust temperature, it is necessary to invoke some additional heating source for the dust in the cone, such as anisotropic emission from the CE, young stars, or even shock heating \citep[see also discussion in][ and Sect.~\ref{additional heating mechanisms}]{Villar-Martin01}. While all these are rather likely to occur, they are not satisfactory explanations in this respect: indeed the dust temperature in the cone, under the assumption of optically thin dust heated only by an isotropic central source, is already too high ($\sim$ 200-300 K at $\sim 0.5$\arcsec, depending on the dimension and composition of grains), to produce a SED declining as steeply as observed towards short wavelengths. A higher T, as could result from an additional heating source, would worsen the problem, independently of the nature of this source.

Therefore, we are led to conclude that a distribution of diffuse dust within the ionizing cone cannot account for the MIR observed SEDs and flux levels.

\subsection{Optically thick dust clouds in the cone}

\begin{figure}[htbp]
\begin{center}
\resizebox{8cm}{!}{\includegraphics*[scale=1.]{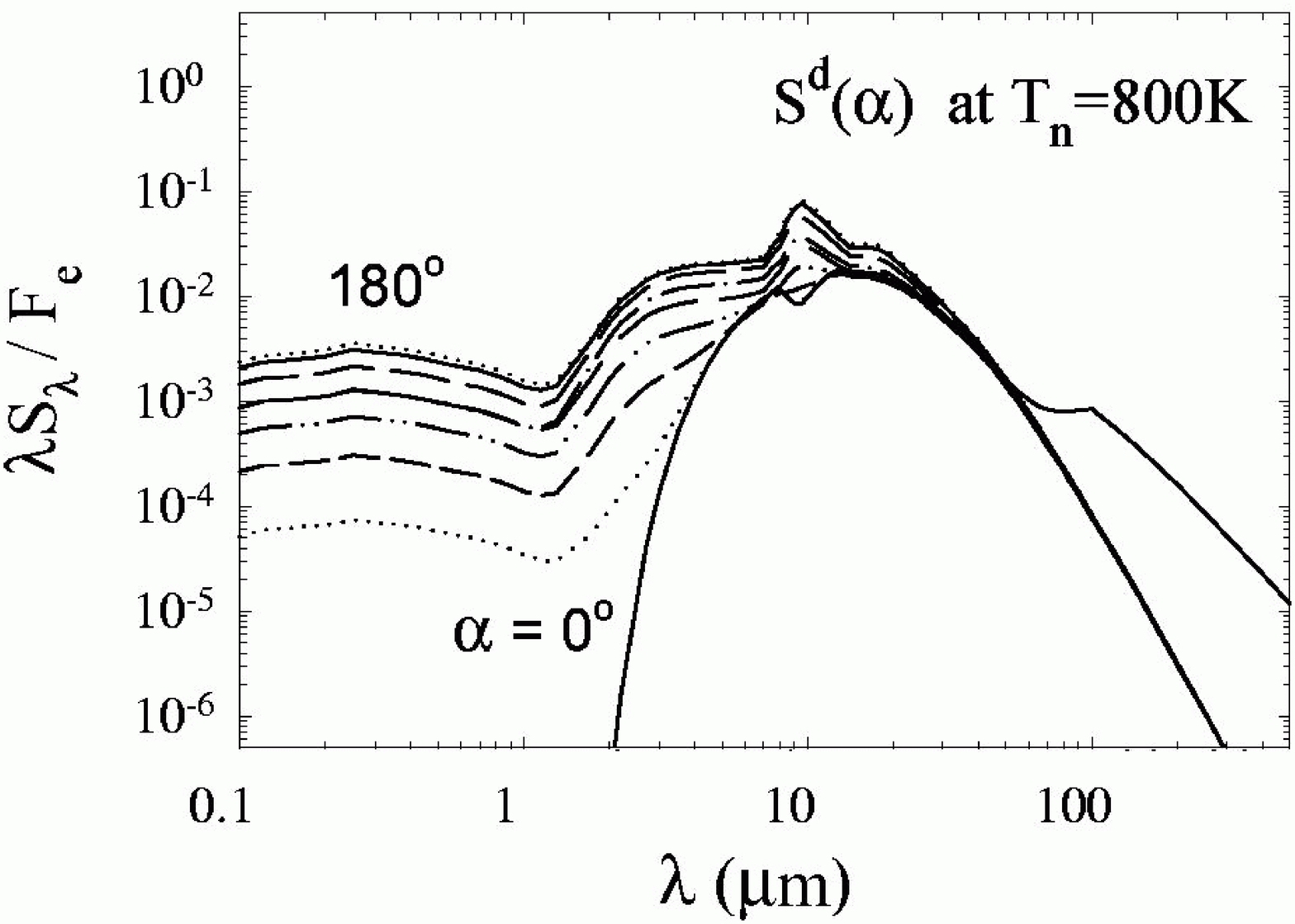}}
\caption{This figure is a partial reproduction of Fig.~2 of \citet{Nenkova02}, reported here for convenience. It shows the predicted SED for an externally illuminated thick ($\tau_V=100$) cloud. The varying parameter $\alpha$ is the angle between the observer and the cloud, as seen from the CE. For $\alpha=0\degr$, the cloud is located between the CE and the observer, and hence, the observer faces the side of the cloud which is not directly illuminated by the CE. In that case, the SED is steeply decreasing towards shorter wavelengths and the silicate feature appears in absorption. For $\alpha=180\degr$, the CE is located between the cloud and the observer, and the observer faces the illuminated side of the cloud. The SED is much flatter than for $\alpha=0\degr$ and the silicate feature appears in emission.  
} 
\label{nenkova}
\end{center}
\end{figure}

The "local" SEDs of the emission displayed in Fig.\,\ref{mapminised} provide another key-element for understanding the MIR emission: the SEDs in the North-East side have a steep decline towards shorter wavelengths and show silicate 9.7 $\mu$m feature in absorption. Both the SED steepness and the depth of the silicate feature increase with distance from the CE. On the contrary, in the South-West side, where the SEDs appear to be flatter, the 9.7\micro~feature is seen in emission. In other words, observations suggest MIR optically thick dust emission in the North-East cone and optically thin emission in the South-West cone. The most natural way to explain this behavior is through the presence of dust within the conical regions, organized in individual optically thick (in the MIR) clouds, and having a small covering factor. Indeed, in such a configuration each cloud would see directly the CE, and would have a much higher dust temperature on its side facing the nucleus than on its opposite side. As a consequence, the overall emission from each individual cloud would be highly anisotropic. In the North-East cone (approaching us) we observe the clouds mostly from their colder surface, while the contrary holds true in the South-West cone (receding from us). Therefore, we expect to see exactly the trend depicted in Fig.\,\ref{nenkova}: the North-East side shows SEDs similar to that labeled $\alpha=0\degr$, while the South-West SEDs are similar to those of the kind labeled $\alpha=180\degr$. 

A picture where the dust is organized in MIR optically thick clouds gives a natural explanation to both the observed emission levels, as well as to the SED shapes, including their slope and silicate feature. The surface brightness of the extended MIR emission can be explained by this model, without invoking additional heating sources. Yet, the observations do not discard either the presence of other sources of heating, which we discuss in the coming section.

\subsection{Possible additional heating mechanisms}
\label{additional heating mechanisms}
In the former section, we discarded a picture where diffuse dust inside the ionizing cone, heated by the CE, was responsible for the observed MIR emission. On the contrary, the local SEDs between 7 and 20\micro~strongly favor a model in which optically thick dusty clouds, with small covering factor, produce the MIR emission. In this context, the radiation from the CE is the major source of heating for these clouds, but the possibility and evidence of additional heating mechanisms ought to be discussed here. 

First, the radiation from the CE might be beamed. This would add a continuum source in the direction of the beam enhancing the dust emission along this region. 

A second possibility would be additional heating of the dust by a starburst. But in the case of {NGC\,1068}, the observed starburst ring is too far away (10\arcsec-20\arcsec) from the MIR structures discussed here. An alternative possibility could be the presence of local star formation along the jet (jet induced star formation). Our superposition of the small scale MIR and radio maps (Fig~\,\ref{superpositions maps}) shows however that the radio emission is much more spatially focused than the MIR emission, and consequently, the hypothesis of jet-induced star formation is not well supported by observations. 

Finally, interactions between radio and optical structures generate shocks that can heat the gas to very high temperatures ($\geq 10^6$\,K). This gas then becomes an efficient source of UV and \Xray~photons able to ionize the surrounding material. Among others, an expected effect is extra heating of the dust. There are indeed some arguments in favor of a contribution of heating by shocks. 

-- the subarcsec NIR continuum emission \citep{Rouan98,Marco00} is elongated along an axis reminiscent of the radio emission axis, and particularly, the 3.5\micro~emission shows a change of direction which follows that of the radio jet (see Fig.\,\ref{superpositions maps}.e). 

-- the subarcsec resolution Chandra (0.25-7.50 keV) image published by \citet{Young01} reveals a bright nucleus and bright extended emission up to 5\arcsec~to the North-East, probably coincident with the extended North-East radio lobe observed at 6cm. Moreover, this image shows that the \Xray~emission immediately to the South of the nucleus has a more or less North-South extension, just like the radio emission. Could this \Xray~emission be contributed by the shocked gas? 

-- \citet{Axon98} find signs of interaction between the radio and optical structures. For example, they find that emission lines are split into two velocity components separated by $\sim$1500\,\kms~at the position of the jet axis. Outside this region, the lines are quiescent. Supporting the shock model is the finding by these same authors of an excess of local optical continuum associated with the radio jet. The existence of an excess of UV continuum is supported by the increase in excitation along the jet axis (outside the jet, the gas excitation is much lower). This continuum excess should naturally contribute to the dust heating, as shown by \citet{Villar-Martin01}. Clues that shocks are present also come from the \H2~2.12\micro~and \Bg~high resolution spectroscopy of \citet{Galliano02}. At the position of the North-\Bg-\H2~knot, corresponding to the position of NLR-cloud E, the \H2~emission is double peaked ($\sim$ 400\kms~velocity difference), showing a kinematical perturbation at this location, possibly induced by shocks. \Bg~is also very broad at this location, displaying a FWHM of $\sim$1000\,\kms, against $\sim$200\,\kms~for \H2. Can we explain this difference with shocks? In the absence of entrainment processes, the cores of the clouds impacted by the jet will be accelerated to a velocity $V$ given by:
$$
V\sim V_{s}\sqrt{\frac{n_h}{n_w}}    
$$
where $V_s$ is the velocity propagation of the shock in the hot phase (that can be identified to the jet advance speed) , and $n_h$ and $n_w$ are the densities of the clouds (ionized or molecular) and of the hot phase ($\sim10^7$\,K). This equation comes from applying the balance between the pressure in the shocked inter-cloud medium ($P_h \sim n_h \times V_{s}^2$), and the pressure behind the shock in the cloud \citep[$P_w \sim n_w \times V^2$,][]{Klein94}.  

Molecular clouds will be accelerated to lower velocities than ionized clouds, due to the higher density (the shock finds higher resistance). Comparing the two velocities:
$$
\frac{V_{B_{\gamma}}}{V_{H_2}}=\sqrt{\frac{n_{B_{\gamma}}}{n_{H_2}}}
$$  
Spectroscopy of the \textit{ionized} regions (i.e. \Bg~emitting gas, but not shocked) not interacting with the radio jet implies densities of a few 100\,cm$^{-3}$ \citep{Axon98}. With the molecular gas being $\sim$ 100 times denser, then $V_{Br_{\gamma}} / V_{H_2} \sim \sqrt{100} = 10$. Therefore, the ionized clouds can easily be accelerated to speeds $\sim$10 times higher than the molecular clouds. This would explain the observed difference in the spectral profiles of \H2~and \Bg~in the North-\Bg-\H2~knot. An alternative possibility is that only molecular clouds impacted by slow shocks can survive, while those enduring high velocity shocks are destroyed. In this case, the \H2~emission would come from more quiescent clouds. 

Therefore, it is likely that shocks, witnessed by the kinematics of the molecular material, play a role in the heating of the dust, although this remains to be quantified. 

\subsection{Conclusion}
The contribution of shocks to the heating of dust and molecular material looks plausible and promising to explain detailed kinematical features. However, we find that the main source of extended MIR emission to the North-East and South-West are MIR optically thick dust clouds distributed within the ionizing cone, with small covering factor.

\section{The molecular emission perpendicular to the cone axis}
\label{H2 model}

The material in the plane perpendicular to the ionization cone axis is visible essentially through molecular line emission of the warm/cold material. No conspicuous NIR or MIR continuum emission is detected along the East-West direction. In particular, the strong CO(2-1) and \H2~2.12\micro~emission lines witness the presence of molecular material up to 100\,pc along the East-West direction \citep{Schinnerer00,Galliano02}. The structures revealed through these lines correlate remarkably well. 

The 0.5\arcsec~resolution \H2~2.12\micro~emission line map of \citet{Galliano02} shows that the \H2~emission is distributed in two unequally bright knots, located at 1\arcsec~(70\pc) from the CE, the brightest to the East and the weakest to the West. Each of the two knots displays a North-South extension which even extends further like an arc or spiral. The East-\H2~knot is 3 times more intense than the West-\H2~one. The CO emission line map has a similar aspect but presents a smaller East-knot to West-knot intensity ratio (1.5 for CO against 3 for \H2). The fact that the CO/H2 ratios are different in the East knot and the West knot suggests that extinction plays a role. 

There is another reason to believe that differential extinction is likely to occur in NGC\,1068, from the observational facts reported on hereafter. Several studies \citep{Schinnerer00,Baker00} argued, on the basis of the CO kinematics, about the presence of a warp in the molecular disc surrounding the AGN of {NGC\,1068}. Moreover, the maps at different scales (see Fig.~\ref{superpositions maps}) show that the PA of the main axis of the observed North-South structures change with scale, which supports the idea of a warp. \Xray~observations show that the material towards the central engine is Compton thick, which implies a column density $\geq 10^{24}\rm\,cm^{-2}$. This Compton thick material (which we will call the \Xray~absorber hereafter) is likely to be located at the very central parts of the AGN, and might be e.g. a corona of hot gas such as the atmosphere of the accretion disk \citep{Collin96}. Since the excitation of molecular emission is dominated by the \Xrays~from the CE \citep{Maloney97}, a tilt between the orientations of the \Xray~absorber and the molecular disk, which would naturally occur in the presence of a warp, would break the axisymmetry and hence could significantly affect the observed molecular line intensity distribution.  

\subsection{Differential extinction between the Eastern and Western knots}

If one assumes that both the East and West knots have intrinsically the same intensity, and that the western emission is dimmed by dust extinction only, a column density towards the West-\H2~knot of the order of $N_{West}=0.5\,10^{22} \rm cm^{-2}$ (for $A_K/A_V=0.11$ and $A_V/N_H=0.19\,10^{-22}\rm\,mag\,(nuclei\,cm^{-2})^{-1}$) is required. This is of the order of the column density through typical molecular clouds in our Galaxy. Dust extinction affects the \H2~line photons, but do not affect the CO line photons at 1.3\,mm. Still, both CO and \H2~photons can be absorbed by molecules in intervening clouds. Large and dense molecular clouds can in fact screen completely the \H2~and CO photons. The estimation of the optical depth through an $N_H=10^{22}\,\mbox{cm}^{-2}$ cloud with $n_{CO}/n_{\mbox{\H2}}=10^{-4}$ results in $\tau>10$ for photons with $\nu(-2\sigma_{turb}) \leq \nu \leq \nu(2\sigma_{turb})$, where $\nu$ is the frequency of the crossing photon, $\sigma_{turb}$ is the velocity dispersion of molecules in the cloud, and  $\nu(2\sigma_{turb})$ is the frequency of the molecular transition for a molecule at radial velocity of $2\sigma_{turb}$. 

In the following, we show quantitatively that a small tilt between the molecular disc and the \Xray~absorber can lead to a notable intensity difference in the \H2 and even, but to a much lesser extent, in the CO emission, between the East and West knots.  

\subsection{The model of \Xray~excitation}
\label{xray excitation}
\begin{figure}[htbp]
\begin{center}
\resizebox{9cm}{!}{\includegraphics*[scale=1.]{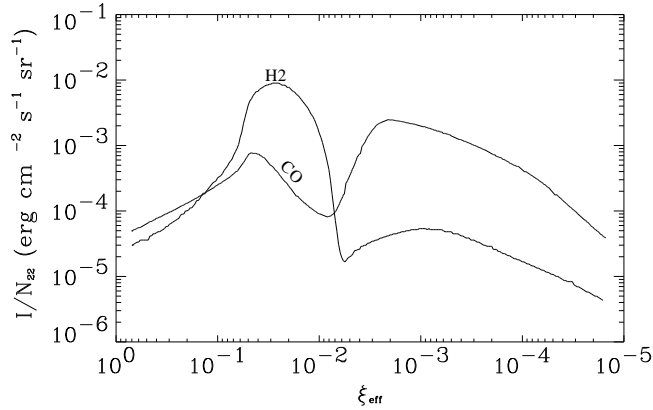}}
\caption{Reproduction of the results for the emergent intensity in the \H2~2.12\micro~emission line and the CO rotational lines for a cloud of column density $10^{22}$\,cm$^{-2}$ \citep{Maloney96}. See Sect.\,\ref{xray excitation} for details.}
\label{Maloney}
\end{center}
\end{figure}

The excitation of the molecular emission is computed using the results of the model of \Xray~irradiated molecular gas developed by \citet{Maloney96}. This model computes the physical and chemical state of dense neutral gas exposed to an intense X-ray flux. We use the numerical predictions of \citet{Maloney96} for a power-law \Xray~source of index $\alpha=-0.7$ ($F_{\nu} \propto \nu^{\alpha}$) and consider molecular clouds of density $n=10^5\,\rm cm^{-3}$ through which the column density is $N=10^{22}\,\rm cm^{-2}$. The most important parameter that controls the physical conditions of the molecular material is the ratio of the attenuated \Xray~flux to density. This parameter is expressed in terms of an effective ionization parameter: 
$$
\xi_{eff}\sim100\,\frac{L_{44}}{\,n_5\,r_{pc}^{2}\,N_{22}^{0.9}}
\label{xieff}
$$
where the 1-100\,keV \Xray~luminosity of the central source is $L_X=10^{44}\,L_{44}$\,\power, the distance to the \Xray~source is $r_{pc}\,\rm pc$, the gas density is $n=n_5\,10^{5}\,\rm cm^{-3}$ and the attenuating column density to the \Xray~source is $N_{att}=10^{22}\,N_{22}\,\rm cm^{-2}$. The predictions of \citet{Maloney96} for the emergent surface brightness of a cloud of density $10^5\rm\,cm^{-3}$ in the 2.12\micro~line and CO rotational lines (all lines together) are reproduced on Fig.\,\ref{Maloney}. 
\begin{figure}[htbp]
\begin{center}
\resizebox{9cm}{!}{\includegraphics*[scale=1.]{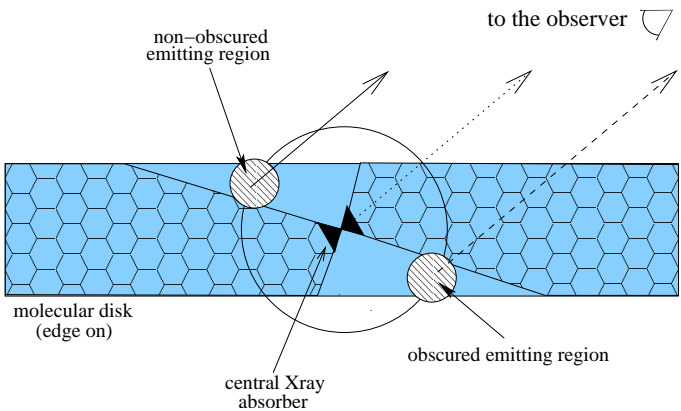}}
\caption{Sketch of a configuration consisting of a molecular disc (grey area) and an \Xray~absorber (black cone) tilted with respect to the disc. See text for details.}   
\label{schema}
\end{center}
\end{figure} 

In the case of {NGC\,1068}, $L_{44}=1$~\citep{Maloney97,Matt97}. Since the maxima of \H2~emission occur at 1\arcsec (70\,pc) from the CE, we know that the value $\xi_{eff}=0.026$ ~(corresponding to the value of $\xi_{eff}$ at the maximum of $I/N_{22}$ for \H2~in Fig.~\ref{Maloney}) corresponds to $r_{pc}$=70. This gives us access to the column density between the CE and the \H2~knots, using the expression of $\xi_{eff}$:
$$
N_{22}=\left(\frac{100\,L_{44}}{\xi_{eff}\,n_5\,d_{pc}^2}\right)^\frac{1}{0.9}=0.75
$$
Therefore, this value is much smaller than the column density of $\geq 10^{24}\rm\,cm^{-2}$ to the CE, deduced from \Xray~observations. This supports the hypothesis that the central \Xray~absorber and the emitting molecular material are not coplanar. Hence, we have developed a new model along this line of arguments, which is presented in the next section.  
 
\subsection{Molecular disc model}
 The configuration sketched in Fig.\,\ref{schema} consists of a molecular disc and of  a central axisymmetrical \Xray~absorber, tilted with respect to the molecular disc. In Fig.\,\ref{schema}, the grey rectangle represents a cut through the edge-on view of the molecular disc. The central \Xray~absorber axis is tilted with respect to the axis of the disc. The region filled with the ``hexagon'' patterns are the regions where the \Xrays~are deeply attenuated, and consequently from which no strong emission can emerge. The circle centered on the CE traces the radius at which, in the absence of a central \Xray~absorber, the value of $\xi_{eff}$ would correspond to the maximum of emission. The regions from which the emission will actually arise are shown with the hatched circles. The observer sees directly the ``non obscured emitting region'', while it sees the ``obscured emitting region'' through the molecular disc. This sketch provides a qualitative explanation to the fact that the molecular emission is offset from the CE, and how it can naturally appear brighter on one side.

\begin{figure*}[htbp]
\begin{center}
\resizebox{13cm}{!}{\includegraphics*[scale=1.]{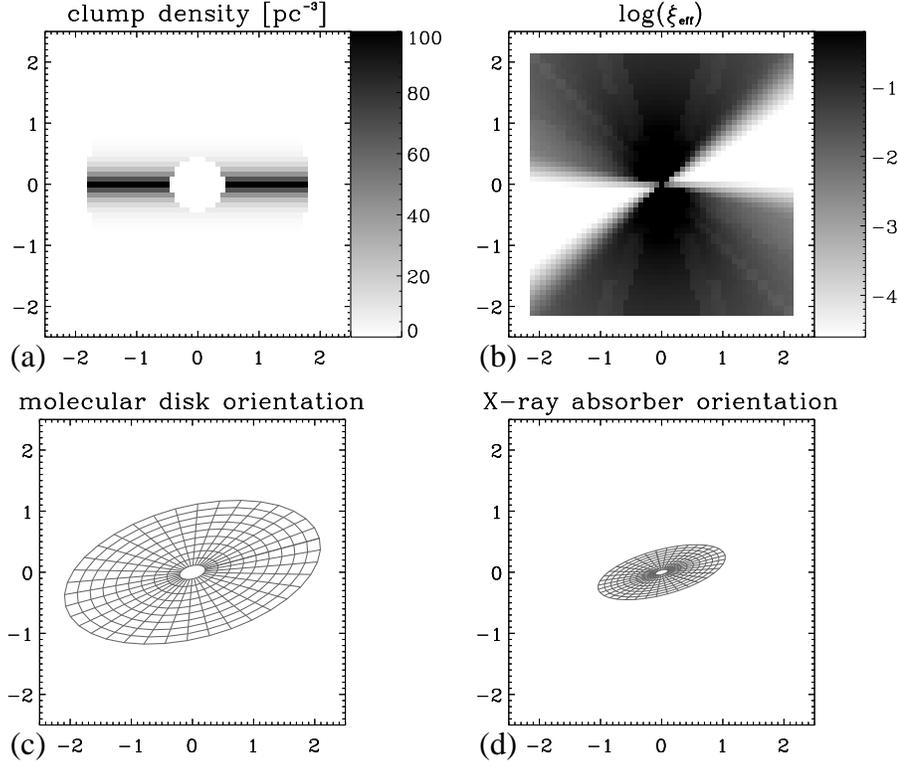}}
\caption{(a) cloud number density distribution for an edge-on view. The image represents the density in the plane perpendicular to the disc and passing through the center. The unit is cloud per pc$^{-3}$. (b) The values of $\xi_{eff}$ for the same plane as in (a). The \Xray~absorber is slightly rotated with respect to the disc. (c) Final orientation of the molecular disc. (d) Final orientation of the \Xray~absorber.}
\label{model result1}
\end{center}
\end{figure*}
\begin{figure*}[htbp]
\begin{center}
\resizebox{13cm}{!}{\includegraphics*[scale=1.]{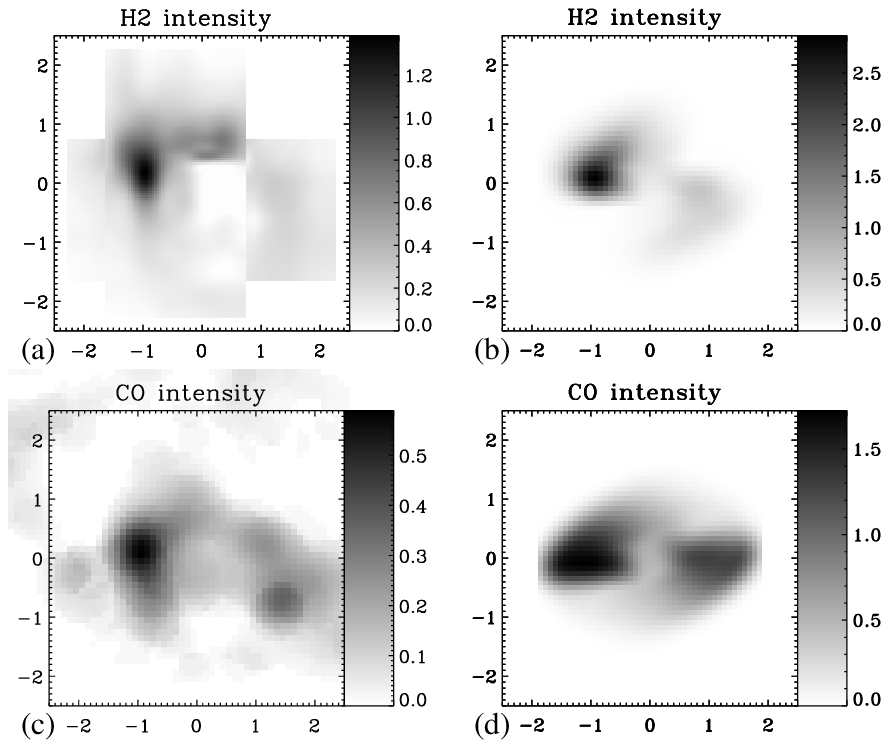}}
\caption{(a) Observed \H2~surface brightness. The unit is $10^{-3}$ \fluxsr. (b) Modeled \H2~surface brightness. (c) Observed CO(2-1)~surface brightness (integrated between -100 and 100\kms). (d) Modeled CO rotation lines~surface brightness. The unit is $10^{-3}$ \fluxsr. In all figures, the x and y axis unit is arcsec.}
\label{model result2}
\end{center}
\end{figure*}

We have built a simple numerical model to simulate such a configuration. It consists of the following two components:

(1) A disc distribution of molecular clouds around the central \Xray~source. All clouds are identical to the densest cloud studied by \citet{Maloney96}, i.e. their density is $10^5\,\rm cm^{-3}$ and their column density is $N=10^{22}\,\rm cm^{-2}$.
Complex cloud number density distributions describing flared or warped discs were computed as well, but a simple distribution (with less free parameters) is sufficient for getting interesting results. In an (x,y,z) reference frame, where z is the axis of the disc, we have chosen the following distribution:
$$
d_c(z)=D_c\,e^{-(|z|/Z)^\lambda}
$$
where $d_c$ is the cloud number density, $D_c$ is the density at z=0, and $(Z,\lambda$) controls the dependence of the density on the height above the equator of the disc. A flat rotation curve was assigned to the disc (plateau at 130\,\kms), in order to estimate the opacity of the clouds in the lines. 
 
(2) A central axisymmetrical \Xray~absorber, for which the equatorial attenuation is very high ($\geq10^{24}\,\rm cm^{-2}$). We have set the azimuthal dependence of its attenuating column density to the following form:
$$
N_{att}(\Theta)=N_{att}(0)\,10^{sin^{\beta}(\Theta)},
$$
where $\Theta$ is the azimuthal angle and $N_{att}(0)$ the equatorial thickness of the \Xray~absorber, that is $N_{att}\geq\,10^{24}\,\rm cm^{-2}$.

\medskip

These two components can be freely and independently rotated in space.

For each cell of a 3D grid we have computed the value of the attenuation towards the \Xray~source, as the sum of the attenuations due to both components. The value of $\xi_{eff}$ in each cell is then computed. 

The attenuation of the emission line photons on their way to the observer is computed taking into account two effects. (1) the dust extinction for \H2~only, (2) the high opacity of some intervening molecular clouds in both \H2~and CO lines (see Appendix~\ref{proba}).

\subsection{Results} 

Fig.\,\ref{model result1} and Fig.\,\ref{model result2} display the results of an illustrating model for a seeing of 0.4\arcsec. 

In Fig.\,\ref{model result1}, the top two figures (a) and (b) present an edge-on view of the disc. For a plane perpendicular to the disc, passing through the center, image (a) shows the cloud density distribution, and image (b) the values of $\xi_{eff}$ for an \Xray~absorber tilted with respect to the molecular disc by 15\degr~in the direction of the x axis. The subsequent simulated images (c,d,f) are not seen edge-on anymore. Images (c) and (d) respectively show the actual orientations of the chosen model for the molecular disc and the \Xray~absorber.

Fig.\,\ref{model result2} compares model predictions with observations:
(a) displays the \H2~emission map and (b) the model prediction for the \H2~emission. (c) displays the CO(2-1)~emission map and (d) the model prediction for the CO (all lines)~emission.

The parameter values corresponding to this model are: $D_c=100\rm\,pc^{-3}$, $Z=20\rm\,pc$, $\lambda=1.15$, $N_{att}(0)=10^{26}\rm\,cm^{-2}$, $\beta=1.5$, $\sigma_{cin}=100\,$\kms~and $\sigma_{turb}=20\,$\kms

The orientation of the two components is illustrated by the ellipses of Fig.\,\ref{model result1} (c) and (d). 

This two component model reproduces quite well the most characteristic features of the observations of the following \H2~2.12\micro~and CO emission: The distance of the observed emission knots to the CE, the extended arc or spiral structure, the intensity ratios between the East and West knots in \H2~and CO (3.2 for \H2 and 1.25 for CO) and is consistent with the Compton thickness of the absorber along the line-of-sight. Yet, the surface brightnesses of the simulated emission knots are too large compared to the observations. Choosing a smaller cloud number density lowers the intensity, but lowers as well the intensity difference between the East and West knots. This suggests that there is probably a difference in the amount of material between the East and the West knots. If we consider such a difference, we can then fit more easily the surface brightnesses and the intensity (all lines together) ratios at the same time. Still, in that case a configuration with selective extinction is required. 

Another point is that, for the chosen model, the predicted CO emission knots are more extended than observed. This can be resolved by adjusting the dependance of the cloud density distribution on the distance to the central engine. The fact that we predicted the total CO rotational emission intensity explains part of this discrepancy. We also point out that, because CO is still emitted at lower $\xi_{eff}$ values than \H2, in the configuration we present, the \H2 and the CO peaks do not fall exactly at the same position, the CO emission being slightly rotated counter-clockwise with respect to the \H2~emission. 

An interesting observational feature, which is not reproduced by this model, is the increase of CO emission towards the South in the western knot, a feature which is not observed for the \H2~emission. In the kinematical configuration proposed by \citet{Galliano02} to fit the \H2~\textbf{line profiles}, both a rotating \textbf{and} an outflowing kinematical components had to be introduced. One notices that it is precisely in the South-West and North-East regions that the projected velocities of both components have the greatest difference (where the profiles are the most asymmetric). In the western region, the outflowing component would be more absorbed by the rotating disc to the North (the projected velocities tend to be more equal), than to the South. Hence, this would qualitatively explain a rise of CO emission to the South of the western knot.

\section{Summary and conclusion}
\label{conclusion}
We have discussed in this paper several features of the central 100\,pc region around the CE in the AGN of {NGC\,1068}. 

The second section was dedicated to revisiting the relative positioning of maps at different wavelengths. We provide a new estimate of the absolute positioning of the unresolved peak in the K-band. After compiling the most important observations published, and after elucidating the apparent disagreement between different results, we could register with a high precision the radio, NIR, optical and UV continuum maps, as well as the submillimiter, NIR, optical and UV emission line maps. We produced a series of map overlays, which constitute an essential tool for giving a general view of the complexity of the central region of {NGC\,1068}. These superpositions show that the MIR is very well correlated with the [OIII] emitting NLR clouds, revealing the survival of dust grains in the UV-irradiated ionization cone. The final registration of the small scale high resolution radio maps with respect to the HST optical maps show a surprising anti-correlation: no NLR cloud is found along the radio jet trajectory. This suggests a picture where the radio jet sweeps away material within and along the axis of the ionizing cone. 

The third section deals with the modeling of the NIR-MIR unresolved peak emission in {NGC\,1068}. First, we could show that the SED fitting approach is somewhat limited since several acceptable combinations of torus geometries and dust parameters are compatible with the observed IR SED of {NGC\,1068}. We conclude that no strong constraint on the geometry of the torus can be set using this approach, while its extension and thickness can to some extent be estimated: the torus in {NGC\,1068} is smaller than 10-20\,pc and moderately thick in the MIR. 

Then, a discussion is provided about the origin of the observed extended MIR emission, mainly along the axis of the ionizing cone. We find that the most natural way to match the observations is to consider individual optically thick clouds, distributed in the NLR and having a small covering factor. An alternative picture, consisting in diffuse dust distributed within the cone, is confidently discarded. Possible additional sources of heating are discussed, and we show in particular that the likely presence of shocks can explain some of the observed kinematical features such as the difference between the FWHM of the molecular emission lines and of the ionized gas emission lines, in the North-\Bg-\H2~knot in particular. 

The last section presents a model of the excitation and distribution of the \H2~and CO line emission along the East-West direction (perpendicular to the ionizing cone axis). We argue that the observed intensity differences between the eastern and the western regions, with respect to the CE, can be partly due to selective extinction produced by a slight tilt between the orientations of the Compton thick \Xray~absorber in the very central region of the AGN and the molecular material distribution at larger scales.

\begin{acknowledgements}
We would like to thank Daigo Tomono, Linda Tacconi and Eva Schinnerer, who kindly provided their data in the MIR and in the CO emission line respectively. We also aknowledge interesting discussions with Laura Silva, about the extended MIR emission in {NGC\,1068}. Finally, we are gratefully indebted to the referee J. Gallimore for valuable comments. E.G. aknowledges the support of an ESO studentship and of the french ministery of Foreign Affairs, as a cooperant.
\end{acknowledgements}

\appendix
\section{Attenuation of \H2~and CO photons by intervening molecular clouds}
\label{proba}
We computed that a molecular cloud of column density $N=10^{22}\,\mbox{cm}^{-2}$~is completely opaque ($\tau\geq10$) to CO and \H2~photons, if the radial velocity difference between the emitting molecule and the molecular cloud is $\leq2\sigma_{turb}$, where $\sigma_{turb}$ is the velocity dispersion inside the cloud (of the order of 10-20\kms).  Let us estimate the corresponding attenuation effect.

We consider two cells of the grid aligned along the line of sight, $cell_a$ and $cell_b$. $Cell_a$ contains one emitting cloud called $cloud_a$ and $cell_b$ contains one non-emitting cloud called $cloud_b$. $Cell_b$ is located between $cell_a$ and the observer. 
We can estimate a cross section $S_{eff}$ of radius $r_{eff}$ for $cloud_a$, so that we can consider that, when the center of $cloud_b$ lies inside $S_{eff}$, $cloud_a$ is completely covered, and when the center of $cloud_b$ lies outside, then $cloud_a$ is completely uncovered. The quantity $r_{eff}$ is such that the overestimation in covering when the center of $cloud_b$ does not coincide with the center of $cloud_a$ (but still lies within $S_{eff}$) is balanced by the underestimation in covering when the center of $cloud_b$ is outside $S_{eff}$ (but still covers part of $cloud_a$). We find $r_{eff}$=0.89$r_{cloud}$. 

Then we can simply compute the absorption in terms of probability for the center of $cloud_b$ to lie within the cross section of $cloud_a$ in the area of a cell $S_{cell}$. This prbability is $P_1=S_{eff}/S_{cell}$. In P1$\times$100\%~of the cases, $cloud_b$ can be considered to be hiding completely $cloud_a$, and in (1-P1)$\times$100\%~of the cases, $cloud_b$ does not hide $cloud_a$ at all. 

We can define as well $V_{eff}$ (the analogous to $r_{eff}$ in the velocity space) such that, if $-V_{eff} \leq V_{a}$-$V_{b} \leq V_{eff}$, we can consider that the photon, in average will be absorbed, where $V_{a}$ and $V_{b}$ are the respective velocities of $cloud_a$ and $cloud_b$ along the line of sight. We find $V_{eff}=1.2\sigma_{turb}$. 

Let $\sigma_{cin}$ be the velocity dispersion between the cloud velocities. If $\Delta V=abs(V_a-V_b)$ is the difference between the radial velocities of $cloud_a$ and $cloud_b$, then the probability for $cloud_b$ to have the same velocity as $cloud_a$ is: $P_2=V_{eff}(\sigma_{cin}-\Delta V)/\sigma_{cin}^2$. 

Finally, in $(P1 \times P2)$\%~of the cases, the photon emitted by $cloud_a$ will be absorbed by $cloud_b$. Thus, we approximate the attenuation by molecules in intervening clouds, by a factor $(1-P_1 \times P_2)$ per cloud in the cells located along the line of sight to the observer.

\end{document}